\documentclass[twocolumn,longbibliography,superscriptaddress,nofootinbib,bibnotes,notitlepage]{revtex4-2}
\usepackage{amsmath}
\usepackage{color}
\definecolor{red}{rgb}{0.75,0,0}
\definecolor{blue}{rgb}{0,0,0.75}
\definecolor{green}{rgb}{0,0.5,0}
\usepackage{amsmath}
\usepackage{amssymb}
\usepackage{bm}
\usepackage{xcolor}
\usepackage{graphicx}
\usepackage{datetime}
\longdate

\makeatletter
\def\maketitle{
	\@author@finish
	\title@column\titleblock@produce
	\suppressfloats[t]}
\makeatother

\def\be{\begin{equation}}
\def\ee{\end{equation}}
\def\bea{\begin{eqnarray}}
\def\eea{\end{eqnarray}}

\def\besub{\begin{subequations}}
\def\eesub{\end{subequations}}

\def\bwd{\begin{widetext}}
\def\ewd{\end{widetext}}

\newcommand{\bsf}[1]{\textsf{\textbf{#1}}}

\newcommand{\qq}{\begin{eqnarray}}
\newcommand{\qqq}{\end{eqnarray}}

\begin{document}
\title{Chirality and odd mechanics in active columnar phases}

\author{S. J. Kole}
\email{sjk202@cam.ac.uk}
\affiliation{Centre for Condensed Matter Theory, Department of Physics, Indian Institute of Science, Bangalore 560 012, India}
\affiliation{INI, University of Cambridge, 20 Clarkson Rd, CB3 0EH Cambridge,United Kingdom}
\affiliation{DAMTP, Centre for Mathematical Sciences, University of Cambridge, Wilberforce Rd, CB3 0WA Cambridge, United
Kingdom}

\author{Gareth P. Alexander}
\email{G.P.Alexander@warwick.ac.uk}
\affiliation{Department of Physics, Gibbet Hill Road, University of Warwick, Coventry CV4 7AL, United Kingdom}

\author{Ananyo Maitra}
\email{nyomaitra07@gmail.com}
\affiliation{{Laboratoire de Physique Th\'eorique et Mod\'elisation, CNRS UMR 8089,
		CY Cergy Paris Universit\'e, F-95032 Cergy-Pontoise Cedex, France}}
\affiliation{Sorbonne Universit\'{e} and CNRS, Laboratoire Jean Perrin, F-75005, Paris, France}

\author{Sriram Ramaswamy}
\email{sriram@iisc.ac.in}
\affiliation{Centre for Condensed Matter Theory, Department of Physics, Indian Institute of Science, Bangalore 560 012, India}
\affiliation{International Centre for Theoretical Sciences, Tata Institute of Fundamental Research, Bangalore 560 089 India}

\date{\today}

\begin{abstract}
Chiral active materials display odd dynamical effects in both their elastic and viscous responses. We show that the most symmetric mesophase with two-dimensional odd elasticity in three dimensions is chiral, polar and columnar, with two-dimensional translational order in the plane perpendicular to the columns and no elastic restoring force for their relative sliding. We derive its hydrodynamic equations from those of a chiral active variant of model H. The most striking prediction of the odd dynamics is two distinct types of column oscillation whose frequencies do not vanish at zero wavenumber. In addition, activity leads to a buckling instability coming from the generic force-dipole active stress analogous to the mechanical Helfrich-Hurault instability in passive materials, while the chiral torque-dipole active stress fundamentally modifies the instability by the selection of helical column undulations.
\end{abstract}

\maketitle

Living matter continually converts chemical energy into work. A description of mechanics and statistics of such microscopically driven materials must either explicitly account for this chemistry \cite{Markovich, Zwickerrev} or introduce a parameter that breaks time-reversal symmetry at the microscopic level \cite{SRJSTAT, LPDJSTAT, RMP, SRrev, Prost_nat, Salbreux_PhysRep}. This latter description, termed active matter, has been used to describe the novel dynamics, mechanics and statistics of a plethora of broken-symmetry phases~\cite{Tap_smec,Tap_chol, julicher2022broken,Ano_sol, adar_joanny, Ano_hex, Deboo2,SJ_chiral_layered}. An active mechanical effect that is the subject of much current attention is ``odd elasticity''~\cite{Vincenzo_rev, Deboo2, Ano_hex} in chiral \cite{Lubensky_Kamien} two-dimensional solids, recently highlighted in a setting related to developmental biology~\cite{Mietke} though anticipated in part in early studies on rotating Rayleigh-B\'{e}nard convection \cite{Echebarria1, Echebarria2}. 
A key feature is the existence of elastic oscillations in a regime in which mechanical inertia is manifestly negligible. This odd dynamics results from a ratio of elastic and viscous coefficients and, in fact, can arise in two distinct ways: as a ratio of odd elasticity to even viscosity or of even elasticity to odd viscosity. In this article we show how odd coefficients emerge naturally through spontaneous translation symmetry-breaking in an active system, in concert with two discrete asymmetries: \textit{chirality}, and \textit{polarity}. We will do this in the spirit of the classical treatments \cite{RamakrishnanTG, Ebner, Archer, LowenPFC} that yield the elastic properties of a solid as a density-wave in a fluid. This requires augmenting model H \cite{HalpHohen} to include active processes \cite{CatesH1, CatesH2}, chirality \cite{SJ_chiral_layered} and polar symmetry breaking. The three-dimensional odd viscosity that then naturally arises also plays an important role in our treatment. 

The long-range effects of active stresses emerge through fluid flow, whereas odd elasticity is a property of solids in two or more dimensions. Odd dynamics also arises in three-dimensional chiral active mesophases with spatially modulated order~\cite{SJ_chiral_layered,Echebarria1}, where the elasticity of the spatial modulation couples to Stokesian hydrodynamics along any fluid directions. We present a study of the most symmetric system---a polar and chiral active columnar phase~\cite{chandrasekhar1977liquid,prost1980liquid,ramaswamy1983breakdown,deGen} with two-dimensional translational order and one fluid direction---possessing odd elasticity in three dimensions. Our article is the first exploration of columnar materials in active systems, despite the abundance of filamentous assemblies in soft matter and biology \cite{barberi2021local,Lubke,grason2015colloquium,atkinson2021mechanics}. Not only is chirality the rule in biological matter, the monomeric units of most biopolymers are non-centrosymmetric. In particular, the microtubule bundles comprising the axon in nerve cell are a realisation of active, columnar, chiral, macroscopically polar matter \cite{rao2018polarity,ahmadi2006hydrodynamics}. Columnar packings of DNA \cite{barberi2021local,livolant_Nature, livolant_column}, if rendered active by transcription, are another candidate.  

Our main result is the prediction of elastic oscillations in chiral and polar active columnar phases in an inertialess regime. {Despite columnar materials breaking three-dimensional rotation symmetry and two-dimensional translation symmetry, the slow dynamics of the structure is governed by the elastic displacement of the columns. This comes from a combination of elastic forces and Stokesian friction coming from the conserved momentum density.}
{In all systems with translational order, {the low Reynolds number dynamics} of the displacement field reads $\dot{\bf u}_{\bf q}={\bsf M}_{\bf q}\cdot{\bf F}_{\bf q}$. For modes with wavenumber $q$, the mobility ${\bsf M}_{\bf q}$ scales as $1/q^2$, {due to the long-range nature of the Stokes flow} while the elastic force density ${\bf F}_{\bf q}$ scales as $\sim q^2$, yielding a characteristic relaxation rate independent of the magnitude of the wavevector $q=|{\bf q}|$ for most directions.} {However, in equilibrium lamellar or columnar systems, there is no elastic response for material deformations with wavevector in liquid directions.}
In {analogy} with the lamellar case \cite{Tap_smec, Tap_chol, SJ_chiral_layered, Chen_Toner}, activity creates a column tension, and hence a nonvanishing $q=0$ response, even for ${\bf q}$ purely {along the columns}. In contrast to the lamellar case, active columnar materials display an emergent non-reciprocal dynamics of ${\bf u}_\perp \equiv (u_x,u_y)$, a consequence of the odd elastic and viscous response of chiral polar systems. Each embodies non-reciprocity individually, through ${\bf F}_{\bf q}$ and ${\bsf M}_{\bf q}$ respectively; the effective non-reciprocal dynamics, in the form of oscillatory modes due to the two components of the displacement field behaving like a position-momentum pair, emerges by combining the non-reciprocal part of one with the reciprocal part of the other. The two ways of doing this---${{\bsf M}_{\bf q}}_{\text{even}}\cdot{{\bf F}_{\bf q}}_{\text{odd}}$ and ${{\bsf M}_{\bf q}}_{\text{odd}}\cdot{{\bf F}_{\bf q}}_{\text{even}}$---have physically distinct dynamical signatures. Not only the ``even-even'' but also the ``odd-odd'' combinations contribute to the reciprocal response. Importantly, because of the long-range nature of the Stokesian dynamics, the oscillation frequency of this mode does not vanish in the limit of zero wave\textit{number}. The frequency and the very existence of this odd collective oscillation, however, does depend on the angle between ${\bf q}$ and the column axis. Indeed, the three-dimensional character of the columnar material is essential: incompressibility forbids this mode for in-plane perturbations. 

Our additional predictions include a fundamental buckling instability driven by an apolar (force dipole) active stress in common with the lamellar case~\cite{SJ_chiral_layered}; in sharp contrast, the effect of chiral (torque dipole) active stresses is to remodel the structure. We now present in detail the theory from which these results follow.

\section{A polarisable chiral active suspension}

We consider a bulk, three-dimensional, two-component system with three-dimensional number density $\psi$ of active polar and chiral units and momentum density ${\bf g} = \rho{\bf v}$, as functions of position ${\bf r}=(x,y,z)$ and time $t$, with overall incompressibility: a constant total mass density $\rho$ and $\nabla \cdot {\bf v} = 0$ for the joint velocity field ${\bf v}$. The degree of vectorial alignment of the particles is accounted for by the polar order parameter field ${\bf P}$. $\psi$ obeys the conserving dynamics  
\begin{equation}
	\label{psieq3d1}
	\partial_t\psi=-\nabla \cdot({\psi}{\bf v})+M\nabla^2\frac{\delta {F}}{\delta\psi}, 
\end{equation}
where $F$ is the free energy that would control the dynamics in the absence of activity; we discuss its form after constructing the dynamical equations for the velocity and the polarisation fields. 

We consider systems in which the Reynolds number is small at the scales of interest, as is the case for typical microbial and soft matter systems. In this limit, the velocity field is governed by the Stokes equation, with viscous forces balancing active and passive forces in the suspension:
\begin{equation}
	\label{vel3dpsi1}
	\nabla \cdot (\bm{\eta} \nabla {\bf v})= -\nabla \psi \frac{\delta F}{\delta \psi}+\nabla \Pi-\nabla \cdot \bm{\sigma}^a,
\end{equation}
where $\bm{\eta}$ is the viscosity tensor whose general form we will discuss in later sections, $\Pi$ is the pressure that enforces the incompressibility constraint $\nabla\cdot{\bf v}=0$, and $\bm{\sigma}^a$ is the active stress. We ignore additional passive force densities involving ${\bf P}$ and $\delta F/\delta{\bf P}$ \cite{StarkLubensky} whose effects in the polar columnar phase are subdominant to the active terms that we consider. 

The expression of active stress in terms of $\psi$ and ${\bf P}$ has three parts, all leading to terms at the same order in gradients in the polar columnar phase:
\begin{equation}
	\label{strs1}
	\sigma^a_{ij}=-\zeta_H\partial_i\psi\partial_j\psi+\zeta_{pa}P_iP_j-\bar{\zeta}_{pc}(P_l\partial_k\psi\epsilon_{ikl}\partial_j\psi)^S ,
\end{equation}
where the superscript $S$ denotes symmetrisation on the free indices, and $\epsilon_{ijk}$ is the Levi-Civita tensor. The first term is familiar from active model H \cite{CatesH1, CatesH2, SJ_chiral_layered, FinlayScr}. The second is the standard active stress for liquid crystals \cite{Simha_Ramaswamy, Hatwalne, RMP, Toner_Tu_Ramaswamy}. The third, with coefficient $\bar{\zeta}_{pc}$, requires both chirality and polarity and is fundamentally biaxial. A force density with the symmetry of $\bar{\zeta}_{pc}$ has not hitherto been examined in active systems.

The polarisation equation, ignoring advection by hydrodynamic flow and motility, is \cite{StarkLubensky}:
\begin{equation}
	\label{poleq}
	\partial_t{\bf P} -(\bm{\Omega}-\lambda \bsf{A})\cdot {\bf P}=-\Gamma_P\frac{\delta F}{\delta {\bf P}},
\end{equation}
where $\bm{\Omega}$ and ${\bsf A}$ are respectively the antisymmetric and symmetric parts of the velocity gradient tensor. 

To complete the description of the polarizable and chiral active fluid, we need to specify the free energy $F$. To do this, we first separate~\cite{Heinonen} $\psi$ into the small-wavenumber part $\psi_0$ of the active-particle concentration, and $\psi_1$ which is modulated on average in the columnar phase. The former plays no role in the hydrodynamics of column displacements and will not be discussed further. We then work with a free energy $F=\int_{\bf r}[f_P+f_{\psi_1, P}]$ where $f_P=(\alpha_P/2)P^2+(\beta_P/4)P^4+(K_P/2)(\nabla {\bf P})^2$ and \begin{multline}
	\label{free}
	f_{\psi_1, P} = \frac{\alpha}{2}\psi_1^2 + \frac{\beta}{4}\psi_1^4 + {C_\parallel \over 2} \bigl( \hat{{\bf P}}\cdot\nabla\psi_1 \bigr)^2 \\ + \frac{C_{\perp}}{2} \bigl[ ({\bf I} - \hat{{\bf P}} \hat{{\bf P}}): \nabla \nabla \psi_1 + q_s^2 \psi_1 \bigr]^2.
\end{multline}
This concludes our construction of the model of a fluid containing chiral and polar active elements. We use this dynamical description in the next sections to obtain a theory of polar, chiral, active columnar materials in which odd elasticity emerges spontaneously.

\begin{figure*}[t]
	\centering
	\includegraphics[width=\linewidth]{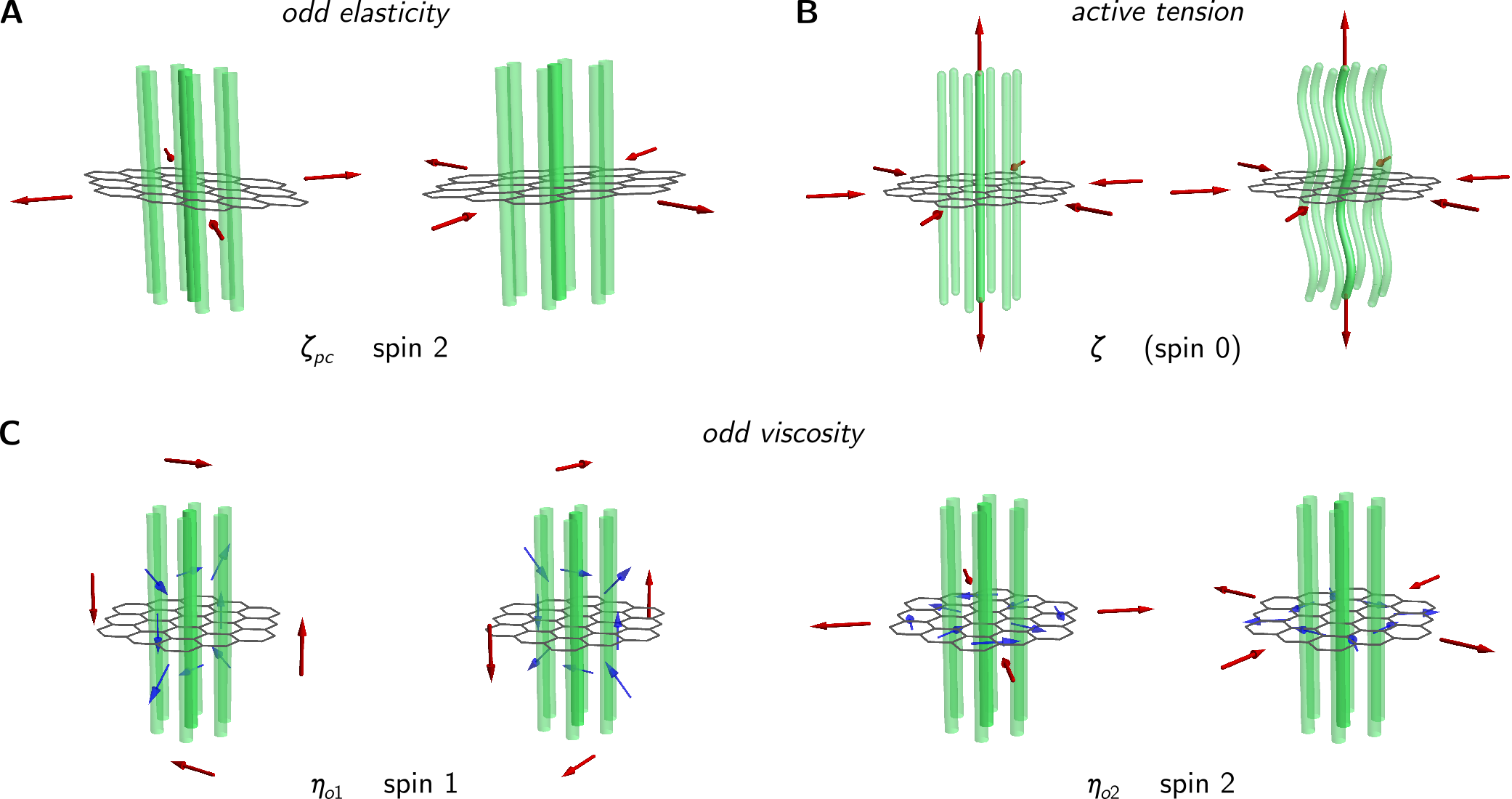}
	\caption{Schematic illustration of stresses (indicated by red arrows) in an active columnar phase; columns are shown as green tubes and the hexagonal arrangement by a centre-plane mesh. (A) There is a single odd elastic modulus in polar chiral materials ($\zeta_{pc}$) coupling deviatoric elastic strains to in-plane deviatoric stresses, with a twist between polarisations; these stresses and strains carry spin $2$. The elastic strains are visualised by the deformation to the regular hexagonal lattice. (B) The regular dipolar activity ($\zeta$) contributes an active tension and precipitates a Helfrich-Hurault-like undulational instability that is isotropic (spin $0$) with respect to the elastic, in-plane directions. (C) There are two odd viscous coefficients ($\eta_{o1}$ and $\eta_{o2}$) associated to twisted couplings between deviatoric shear flows in the spin $1$ and spin $2$ irreps and corresponding stresses. The structure of the fluid velocity in these couplings is indicated by blue arrows.}
	\label{fig:ColumnarSchematic}
\end{figure*}

\section{Breaking polar and translation symmetry: Generation of odd elastic force density} \label{sec:genodd}

While the active fluid described above contains motile units, it \emph{cannot} form a polar nematic phase \cite{Clark2020}. This is because a uniaxial orientational order is necessarily unstable in bulk Stokesian active fluids \cite{Simha_Ramaswamy, RMP, Ano_LRO}. Instead, when $\alpha_P$ and $\alpha$ are \emph{both} negative, the fluid spontaneously breaks both translation and rotation symmetries to form a polar and chiral columnar phase, at least in the absence of activity. The steady-state value of ${\bf P}$ is ${\bf P}_0=\sqrt{|\alpha_P|/\beta_P}\,\hat{{\bf z}} \equiv P_0\,\hat{{\bf z}}$ and that of $\psi_1$ is 
\begin{equation}
	\label{psi1eq}
	\bar{\psi}_1 = \sum_{{\bf G}\in\Lambda^*} \psi_{1,{\bf G}} \,e^{i{\bf G}\cdot{\bf r}} ,
\end{equation}
where ${\bf G}$ are the vectors of the reciprocal lattice $\Lambda^*$ formed by the columnar phase in the plane transverse to $\hat{{\bf z}}$. In the analysis below, we take the fundamental star of a triangular structure with the scale $|{\bf G}| = q_s$ favoured by the free energy and $\psi_{1,{\bf G}} \equiv \psi_{1}^{0} = \sqrt{|\alpha|/\beta}$. Importantly, polar order evades instability \cite{Simha_Ramaswamy, RMP} here only because it is accompanied by \emph{translational} order. Ignoring singularities such as dislocations \cite{julicher2022broken}, we now ascertain the effect of activity on the dynamics of the equilibrium columnar state\footnote{The treatment applies generically to any such state, with our ignorance of the details of the base state buried in the phenomenological coefficients \cite{forster}. Departures from the generic case would require special tuning of coefficients \cite{Cross-Tersauro, Pomeau-Manneville, Ano_sol}.}. Considering broken-symmetry fluctuations about the state $(\bar{{\psi}}_1,{\bf P}_0)$, writing the phase of $\bar{{\psi}}_1$ as ${\bf G}\cdot({\bf r}-{\bf u}_{\perp})$ and ${\bf P}\approx P_0 (\delta{\bf P}_\perp, 1)$, the free energy becomes 
\begin{equation}
	\label{fenergy}
	\begin{split}
		F & = \int_{\bf r} \biggl[ \frac{\lambda}{2} (\text{Tr}\, {\bsf E})^2 + \mu{\bsf E}:{\bsf E} + \frac{K}{2}(\nabla^2{\bf u}_\perp)^2 \\
		& \qquad + \frac{C}{2} (\delta{\bf P}_\perp - \partial_z{\bf u}_\perp)^2 \biggr] ,
	\end{split}
\end{equation}
where $C,\lambda,\mu, K$ are functions of $C_\perp,\, C_\parallel,\,\psi^0_1$ and $q_s$. The nonlinear elastic strain tensor ${\bsf E}$ has components $E_{ij} = \frac{1}{2}(\partial_i u_j + \partial_j u_i - \partial_i u_k \partial_j u_k - \partial_z u_i \partial_z u_j)$, with $i,j,k$ ranging over $x,y$; we retain only its linearized form $E_{ij}=\frac{1}{2}(\partial_iu_j+\partial_ju_i)$ in this article. We will assume, as in equilibrium systems, that $\delta{\bf P}_\perp$ relaxes to $\partial_z{\bf u}_\perp$ in a microscopic time. This assumption holds in the system under consideration provided $\zeta_{pa}/\eta\ll \Gamma_P$ \cite{Ano_sol,Simha_Ramaswamy}. 

Expanding $\psi$ and ${\bf P}$ in terms of ${\bf u}_\perp$ in \eqref{vel3dpsi1} and \eqref{strs1}, we obtain the displacement field-dependent part of the force density whose linearised form is \cite{supp}
\begin{multline}
	\label{elF}
	\boldsymbol{\mathcal{F}}^e=(\bar{\lambda}+\bar{\mu})\nabla_\perp\nabla_\perp\cdot{\bf u}_\perp+\bar{\mu}\nabla_\perp^2{\bf u}_\perp+\zeta\nabla^2{\bf u}_\perp+\zeta_{pc}\nabla_\perp^2\boldsymbol{\epsilon}\cdot{\bf u}_\perp.
\end{multline}
where $\bm{\epsilon}$ is the two-dimensional Levi-Civita tensor and
we have defined the modified Lam\'e coefficients $\bar{\mu}=\mu-\zeta_1$ and $\bar{\lambda}=\lambda-\zeta_2$, which are renormalised by the active terms in \eqref{strs1} with the $\zeta$s being functions of $P_0$, $\psi_1^0$, $q_s$, $\zeta_H$ and $\zeta_{pa}$. The active force density $\zeta\nabla^2{\bf u}_\perp$ has a piece $\propto \partial_z^2{\bf u}_\perp$ (Fig.~\ref{fig:ColumnarSchematic}B) analogous to column tension, that would have been forbidden in equilibrium materials due to a combination of rotation invariance and time-reversal symmetry. This is akin to the activity-induced emergence of a layer tension in lamellar materials \cite{Tap_smec, Tap_chol, Chen_Toner} or an effective elasticity modulus in active nematic elastomers \cite{Ano_sol} or a line tension or surface tension in polymers \cite{Kikuchi} or membranes \cite{Ano_mem, RamTonPro} in active fluids. The chiral force density $\propto\zeta_{pc}$ emerging from the $\bar{\zeta}_{pc}$ term in \eqref{strs1}---exactly the odd elasticity discussed in \emph{two-dimensional} active solids \cite{Deboo2, Ano_hex}---arises naturally in this three-dimensional system through polarity and the spontaneous breaking of translation symmetry, Fig.~\ref{fig:ColumnarSchematic}A. 

While the primary effect of \textit{achiral} active force densities in the columnar materials is analogous to those in active lamellar phases---in the sense that they both endow the system with rigidities that would vanish in equilibrium---the effect of chirality in these two systems is truly distinct. In particular, the two-dimensional odd elastic force density $\propto\zeta_{pc}$ has no analogue in lamellar materials.

\section{Odd viscosity and odd elasticity} \label{sec:oddodd}

We now discuss the general form of the viscosity in a polar chiral columnar material, including its odd contributions. For a uniaxial system with at least sixfold symmetry perpendicular to the preferred axis, the deviatoric part of the velocity gradient tensor is decomposed into irreducible representations (irreps) as $1 \oplus 2 \oplus 2$; the one-dimensional irrep has spin $0$, while the two-dimensional irreps have spin $1$ and spin $2$, respectively. The viscosity tensor maps these irreps of the velocity gradients to a corresponding decomposition of the stress. The normal viscous response is an identity between corresponding irreps, implying three coefficients of viscosity\footnote{A uniaxial system has five viscosities of which the two bulk viscosities play no role in the incompressible limit.} in the even viscous stress: $\eta_{1}\hat{z}_i\hat{z}_jA_{zz}+2\eta A_{ij}+({\eta_3}/{2})(\hat{z}_iA_{jz}+\hat{z}_jA_{iz})$. 

Rotations within the two two-dimensional irreps yields ``odd'' linear mechanical responses. Of course, a rotation requires uniquely defining a vector ${\bf w}_{\perp}$ perpendicular to a given displacement or velocity ${\bf w}$. In three dimensions, this requires chirality in the form of the Levi-Civita tensor $\epsilon_{ijk}$, a preferred axis, say the normal to the $xy$ plane, and a polarity along that axis, hence a unique $\hat{\bf z}$. We can then unambiguously write $w_{\perp i} = \epsilon_{ijz}w_j$. Applying this construction to the linear relation between stress and velocity gradient for any three-dimensional chiral system with a polarity along $\hat{\bf z}$ yields\footnote{Odd viscosities cannot arise in three-dimensional apolar or isotropic chiral systems \cite{Son, Deboo2}.} \cite{supp} 
\begin{equation}
	\label{chivis}
	{\sigma_{ij}}^v= 2\eta A_{ij} + 2\left[\epsilon_{ilz}\left\{{\eta}_{o2}\left(A_{lj}-\delta_{jz} A_{lz}\right)+2{{\eta}_{o1}}\delta_{jz}A_{lz}\right\}\right]^S,
\end{equation}
where ${\eta}_{o1},{\eta}_{o2}$ are the odd viscosities of the spin $1$ and spin $2$ irreps, respectively, and we have set $\eta_1=\eta_3=0$, as these anisotropic contributions to the even viscous response do not qualitatively modify the linearised dynamics of the state of interest. We show a schematic of the two odd viscosities in Fig.~\ref{fig:ColumnarSchematic}C. Though we call ${\eta}_{o1},{\eta}_{o2}$ viscosities, they are not dissipative coefficients in the momentum equation: they break Onsager symmetry and have no relation to noises. Unlike the stress that ultimately resulted in two-dimensional odd elasticity, the odd viscous stress we describe is allowed in \emph{any} chiral material with broken time-reversal and polar anisotropy. \eqref{chivis} implies the odd viscous force density
\begin{multline}
	\label{chivisF}
	\boldsymbol{\mathcal{F}}^v_o = \eta_{o2} \nabla_\perp^2 \bm{\epsilon} \cdot{\bf v}_\perp + {\eta_{o1}} \partial_z \bigl( 2\Omega_{xy} \hat{\bf z} + \partial_z \bm{\epsilon}\cdot{\bf v}_{\perp} +\bm{\epsilon}\cdot\nabla_{\perp} v_z \bigr) , 
\end{multline}
where the contribution $\propto\eta_{o2}$ appears in two-dimensional, chiral active fluids \cite{Deboo1, Vincenzo_rev, Avron} where it can be absorbed into a redefinition of the pressure in an incompressible system. In contrast, the \textit{three}-dimensional incompressibility constraint for our system permits $\eta_{o2}$ to affect bulk flows and column oscillations. The odd viscous force density $\propto\eta_{o1}$, with no analogue in two-dimensional fluids, has an important consequence for the mode structure we discuss in the next section.

We now rationalize the odd elasticity constructed in section \ref{sec:genodd} in a manner analogous to the foregoing treatment of odd viscosities. In a columnar material the component of strain $\partial_z {\bf u}_\perp$ is absorbed by the polarisation fluctuations \cite{deGen} and incurs a cost only at next order in gradients in the form of column-bending elasticity. Leading-order elasticity survives only for the orthogonal strains $\nabla_{\perp} {\bf u}_{\perp}$ which under the local symmetry decompose as $1 \oplus 1 \oplus 2$. The antisymmetric part represents a rigid rotation and does not generate a stress; there are therefore \textit{three} moduli, two even, a bulk and shear modulus, and one odd. We show a schematic of the odd elasticity in Fig.~\ref{fig:ColumnarSchematic}A. However, our approach that naturally yields odd elasticity as a consequence of translation symmetry breaking in active model H establishes it as a new generalised rigidity arising in active, chiral materials. 

Less formally, there are two odd viscosities and one odd elastic modulus because the velocity field is three-dimensional whereas the displacement field in a columnar phase has only two components. In phases with a three-component displacement field, there would be two odd elastic moduli.

\section{Putting everything together: wavenumber-independent odd oscillations in liquid crystals} \label{sec:everything}

The linearised Stokes equation containing both the odd elastic force density discussed in Sec. \ref{sec:genodd} and the odd viscous force density discussed in Sec. \ref{sec:oddodd} is
\begin{equation}
	\eta\nabla^2{\bf v}+\boldsymbol{\mathcal{F}}^v_o=\nabla\Pi-\boldsymbol{\mathcal{F}}^e,
\end{equation}
where $\Pi$ is the pressure enforcing the three-dimensional incompressibility constraint $\nabla\cdot{\bf v}=0$. To leading order in gradients, the dynamics of the displacement field reads $\dot{\bf u}_\perp={\bf v}_\perp$. Fourier transforming and solving for the velocity field yields a dynamical equation for the displacement field in the form $\dot{\bf u}_{\bf q}={\bsf M}_{\bf q}\cdot{\bf F}_{\bf q}$ \cite{supp}, where ${\bf F}_{\bf q}$ arises from $\boldsymbol{\mathcal{F}}^e$ and ${\bsf M}_{\bf q}$ from the viscous force densities with an odd part due to $\boldsymbol{\mathcal{F}}^v_o$.

For much of the rest of the article, we will assume that the magnitudes and signs of these active terms lie in the range in which the columnar phase is linearly stable. Instabilities, when they arise, can do so through both even-even and odd-odd combinations of mobility and force density; a comment about the even-even case $\zeta<0$ is in order here. The achiral active force density in an incompressible columnar system is \emph{exactly} the same as that of an external stress, uniaxial along the columns or, equivalently, isotropic in the plane normal to them even considering the full nonlinear dynamics. This is analogous to active smectics \cite{SJ_chiral_layered}. This implies that as in cholesterics and smectics \cite{SJ_chiral_layered}, $\zeta<0$ leads to a spontaneous Helfrich-Hurault instability \cite{TranRMP} with, however, a degeneracy in the polarisation of the two-dimensional column-undulation field. In a finite system of lateral size $L$, this instability happens beyond a critical active stress $\pi{\sqrt{K\bar{\lambda}}}/{L}$ \cite{supp}. As in lamellar phases \cite{SJ_chiral_layered}, the nonlinear mapping implies that beyond this instability an active \emph{achiral} columnar material assumes the same state as an equilibrium columnar subject to an external stress. 

\begin{figure*}[t]
	\centering
	\includegraphics[width=1.0\linewidth]{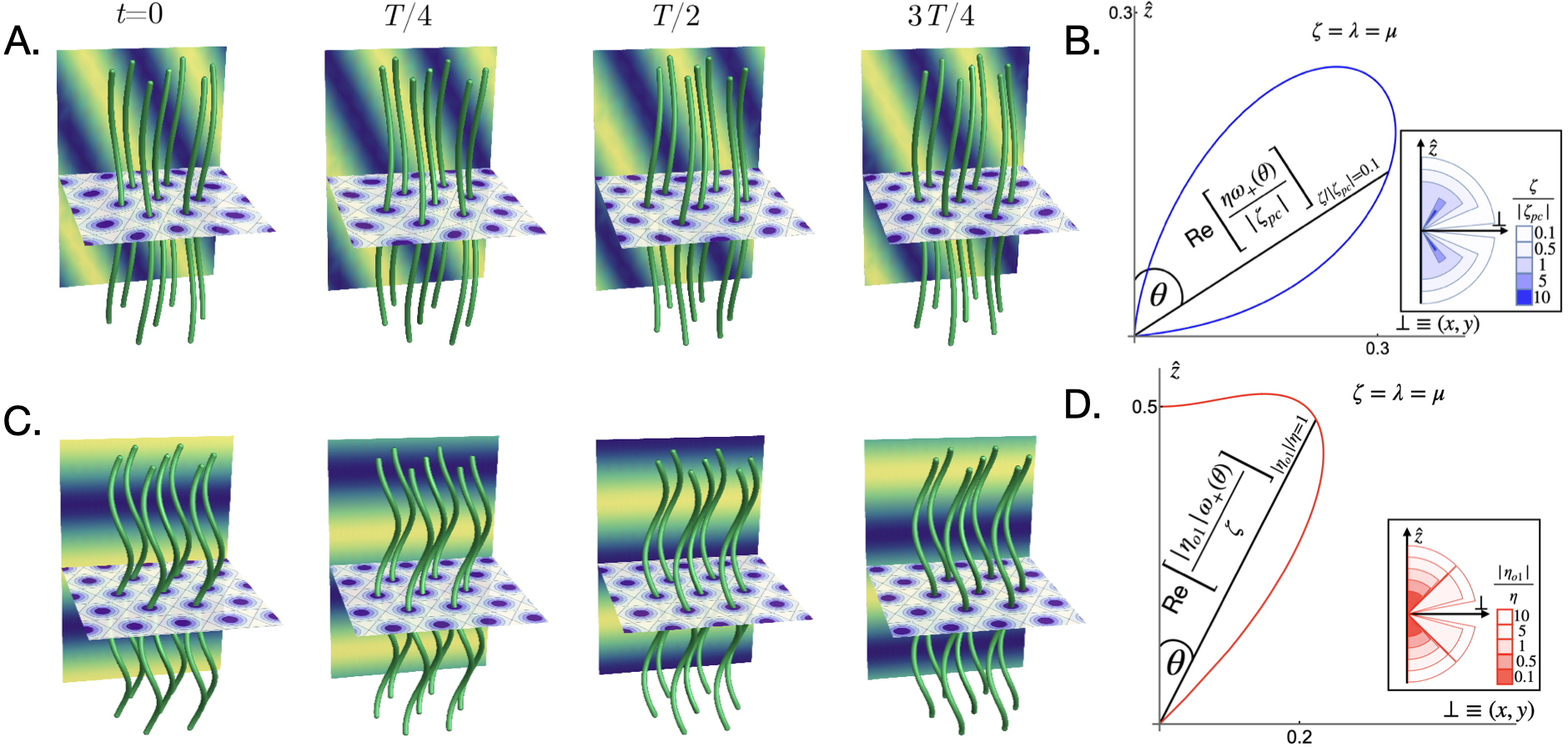}
	\caption{Odd dynamical oscillations in active polar chiral columnar phases. (A) Odd elastic plane wave solution of~\eqref{ulpol} and~\eqref{utpol} due to ${\bsf M}_{{\bf q}_{\text{even}}}\cdot{\bf F}_{{\bf q}_{\text{odd}}}$; here, $\eta_{o1}=\eta_{o2}=0$ and $\zeta_{pc}\neq 0$. The horizontal plane shows a density plot of $\psi$ and the background plane indicates the phase and direction of the plane wave. Columns are shown as green tubes and images are given for every quarter of an oscillation period ($T$). (B) Polar plot of $\text{Re}[\eta\omega(\theta)_+/|\zeta_{pc}|]$---the oscillation frequency, non-dimensionalised by the active timescale $\eta/|\zeta_{pc}|$---of the odd elastic oscillations due to the combination of even mobility and odd force density for $\lambda=\mu=\zeta$. $\theta=0$ corresponds to a perturbation purely along the $\hat{{\bf z}}$ (polarity) direction while $\theta=\pi/2$ corresponds to a perturbation purely along the in-plane, crystalline directions. \textit{Inset}: Colour-coded sectors (with different radii to ensure visibility) showing the range of $\theta$---the angle between the wavevector of perturbation and $\hat{{\bf z}}$---that elicits an oscillatory response for various values of $\zeta/|\zeta_{pc}|$. This demonstrates that the angular range increases with decreasing $\zeta/|\zeta_{pc}|$. While the angular range for $\zeta/|\zeta_{pc}|\leq1$ extends \emph{almost} to $\theta=0$, there is no oscillatory response for a perturbation with wavevector along $\hat{{\bf z}}$. Instead, $\text{Re}[\omega_\pm]\propto\theta^2$ at small $\theta$. (C) Odd oscillations from the interplay of odd mobility and even force density, ${\bsf M}_{{\bf q}_{\text{odd}}}\cdot{\bf F}_{{\bf q}_{\text{even}}}$: plane wave solution of~\eqref{uloddvisc} and~\eqref{utoddvisc} with wavevector purely along the column axis direction. Here, $\eta_{o1}\neq 0$. (D) Polar plot of $\text{Re}[|\eta_{o1}|\omega(\theta)_+/\zeta]$---the oscillation frequency, non-dimensionalised by the timescale $|\eta_{o1}|/\zeta$---of odd oscillations due to the combination of odd mobility and even force density for $\lambda=\mu=\zeta$. Here, $\zeta_{pc}=\eta_{o2}=0$. \textit{Inset}: Colour-coded sectors (with different radii to ensure visibility) showing the range of $\theta$---the angle between the wavevector of perturbation and $\hat{{\bf z}}$---that elicits an oscillatory response for various values of $|\eta_{o1}|/\eta$. Note that the oscillation vanishes for $\theta=\pi/4$ because the mobility due to $\eta_{o1}$ vanishes at this value in incompressible materials as can be seen from the expression of $\nu_o$ below \eqref{utoddvisc}.}
	\label{fig:polar_wave_3d}
\end{figure*}

We now turn to an examination of the effect of odd elasticity and viscosity on the dynamics of active, polar and chiral columnar phases when they are linearly stable. We begin with odd elasticity alone, setting $\eta_{o1},\eta_{o2}=0$. The resulting response of the columnar material to a perturbation with wavevector ${\bf q}=({\bf q}_\perp,q_z)=q(\sin\theta\cos\phi,\sin\theta\sin\phi,\cos\theta)$ has the form $\dot{\bf u}_{\bf q}={{\bsf M}_{\bf q}}_{\text{even}}\cdot({{\bf F}_{\bf q}}_{\text{odd}}+{{\bf F}_{\bf q}}_{\text{even}})$.
Defining its components $u_l={\bf q}_\perp\cdot{\bf u}_{\bf q}/|q_\perp|$ and $u_t=(q_xu_y-q_yu_x)/|q_\perp|$ along and transverse to $q_\perp $ we find that ${{\bf F}_{\bf q}}_{\text{odd}} = \zeta_{pc} q_{\perp}^2 \,\boldsymbol{\epsilon}\cdot{\bf u}_{\bf q}$, couples the $u_l$ and $u_t$ dynamics at zeroth order in wavenumber \cite{supp}: 
\begin{gather}
	\label{ulpol}
	\dot{u}_l = - {(2\bar{\mu}+\bar{\lambda})q_{\perp}^2 q_z^2 + \zeta q^2 q_z^2 \over \eta q^4} u_l - {\zeta_{pc} q_z^2 q_{\perp}^2 \over \eta q^4} u_t , \\
	\label{utpol}
	\dot{u}_t = - {\bar{\mu} q_{\perp}^2 +\zeta q^2 \over \eta q^2} u_t + {\zeta_{pc}q_{\perp}^2 \over \eta q^2} u_l.   
\end{gather}
This Poisson-bracket-like coupling is truly three-dimensional, vanishing for $q_z = 0$ or $q_\perp=0$\footnote{Its equivalent is ruled out in an incompressible, two-dimensional odd gel where, to $\mathcal{O}(q^0)$, $\dot{u}_l$ and the longitudinal velocity vanish. In the present work, however, incompressibility constrains the \textit{three}-dimensional velocity field and, therefore, $\dot{u}_l \neq 0$ at $\mathcal{O}(q^0)$ for a perturbation with nonzero $z$ component.}. As discussed in the Introduction, \eqref{ulpol} and \eqref{utpol} leads to an oscillatory response for large-enough $\zeta_{pc}$, vanishing only for perturbations purely in the plane or purely along $\hat{z}$. Figure \ref{fig:polar_wave_3d}A shows a representative time series of the column distortions associated to this odd oscillation; the dynamics is right-handed for postive $\zeta_{pc}$. The general expression for the eigenfrequency displayed in \cite{supp} is complicated, but its essential features can be seen in its form for $\theta=\pi/4$, 
\begin{equation}
	\label{osceigen}
	\omega_\pm(\theta=\pi/4) = \pm \frac{\mid{\zeta_{pc}}\mid}{2\sqrt{2}\eta}\sqrt{1 - \left(\frac{\bar{\lambda} - 2\zeta}{2\sqrt{2}\zeta_{pc}}\right)^2} - i \frac{6\zeta+4\bar{\mu}+\bar{\lambda}}{8\eta}.
\end{equation} 
\eqref{osceigen} describes an oscillatory mode when $|\zeta_{pc}| > |\bar{\lambda}-2\zeta|/2\sqrt{2}$. For a general angle $\theta\neq 0,\pi/2$ between ${\bf q}$ and $\hat{{\bf z}}$, the response is oscillatory when $|\zeta_{pc}|>|\zeta+\bar{\mu}-(2\bar{\mu}+\bar{\lambda})\cos^2\theta|/2|\cos\theta|$. This implies that when $\zeta_{pc}$ is larger than the other active and elastic terms, a perturbation with a wavevector \emph{almost} fully along $\hat{{\bf z}}$ also leads to an oscillatory response with $\lim_{\theta\to 0}\mathrm{Re}[\omega_\pm]\propto\theta^2$. We display the angular ranges in which we predict an oscillatory response in Fig.~\ref{fig:polar_wave_3d}B for various values of $\zeta_{pc}$.

It is instructive to compare with the dispersive wave induced by non-reciprocity in two-dimensional compressible solids on frictional substrates \cite{Deboo2}, where odd elasticity leads to the dynamics $\dot{u}_l\propto -q_\perp^2 u_t$ and $\dot{u}_t\propto q_\perp^2 u_l$. In active polar and chiral columnar phases, the presence of momentum conservation and three-dimensional incompressibility leads to a radically different mode structure. Momentum conservation induces a long-range interaction, replacing the dispersive wave with an oscillatory mode whose frequency is nonvanishing but non-analytic for $q\to 0$. In the presence of a momentum sink, the damping in the momentum equation would ultimately imply a dispersive wave at small wavenumbers. Indeed, if the columnar material is confined in a geometry which has a finite extent $\ell$ in the $z$ direction, with porous walls such that the fluid can flow out of the system, then one can replace $q_z^2$ with $1/\ell^2$ to obtain the displacement dynamics in \cite{Deboo2} quoted above. Note that the three-dimensional character of the active liquid crystal remains crucial, as the mode involves a $z$-variation of the $z$-component of the velocity field at the scale of the system. 

If instead the confinement is on a scale $\ell$ in the $\perp$ plane, the odd elastic coupling for $q_z \ell \ll 1$ takes the form $\dot{u}_l\propto -(\zeta_{pc}/\eta)\ell^2 q_z^2u_t$ and $\dot{u}_t\propto (\zeta_{pc}/\eta) u_l$. The possibility of active non-dispersive waves \`{a} la \cite{RamTonPro, Ano_mem} is, however, ruled out by the effective damping rates $(\mu+\zeta)/\eta$ and $(2\mu+\lambda+\zeta)\ell^2q_z^2/\eta$ of $u_t$ and $u_l$ respectively. The resulting eigenfrequencies $\omega_+=-i(\zeta+\bar{\mu})/\eta$ and $\omega_-=-i\ell^2q_z^2[(\zeta+\bar{\lambda}+2\bar{\mu})/\eta + \zeta_{pc}^2/\eta(\zeta+\bar{\mu})]$ have no real part. 

We now turn to the dynamics with non-zero odd viscosities and examine their effect on the displacement dynamics focusing on the character of the oscillatory response due to the combination of ${{\bsf M}_{\bf q}}_{\text{odd}}$ and ${{\bf F}_{\bf q}}_{\text{even}}$. The equations of motion for $u_l$ and $u_t$ are \cite{supp}
\begin{align}
	\dot{u}_l & = - \frac{\eta [(2\bar{\mu}+\bar{\lambda})q_{\perp}^2+\zeta q^2]q_z^2}{\Delta q^4} u_l + \frac{\nu_{o}(\bar{\mu}q_{\perp}^2+\zeta q^2)q_z^2}{\Delta q^4} u_t , 
	\label{uloddvisc} \\
	\dot{u}_t & = - \frac{\nu_{o} [(2\bar{\mu}+\bar{\lambda})q_{\perp}^2+\zeta q^2]q_z^2}{\Delta q^4} u_l - \frac{\eta(\bar{\mu}q_{\perp}^2+\zeta q^2)}{\Delta q^2} u_t , 
	\label{utoddvisc}   
\end{align}
with $\nu_o=\eta_{o1}(q_z^2-q_\perp^2)/q^2+\eta_{o2}q_\perp^2/q^2$ and  $\Delta=\eta^2+\nu_o^2q_z^2/q^2$. We highlight a few important features of \eqref{utoddvisc} and \eqref{uloddvisc}. As discussed in the Introduction, odd-odd and even-even couplings of mobility and force density lead to dissipative terms, and odd-even or even-odd to reactive couplings between $u_l$ and $u_t$, as is generically the case in chiral active systems \cite{Ano_hex}. 

Importantly, while the odd force $\propto\zeta_{pc}$ vanishes for perturbations purely along the $z$-direction, the apolar and achiral active force density $\propto\zeta$ \emph{does not}. Of the two odd contributions to the mobility, one $\propto\eta_{o2}$ vanishes for perturbations purely along $\hat{{\bf z}}$, but the other $\propto\eta_{o1}$ does not. In fact, for a perturbation with $\theta=0$ the coupled $u_l-u_t$ dynamics \eqref{uloddvisc} and \eqref{utoddvisc} has eigenfrequencies $\omega_\pm=-\zeta/(\eta\pm i\eta_{o1})$ implying an oscillatory response \emph{even} for perturbations purely along $\hat{{\bf z}}$, Fig.~\ref{fig:polar_wave_3d}C. That is, unlike in Fig.~\ref{fig:polar_wave_3d}B, the eigenfrequency no longer vanishes in the $\theta\to 0$ limit. We emphasise that this is purely due to odd \emph{viscosity} of three-dimensional polar, chiral materials---and its interplay with achiral active line tension---and not odd elasticity. Significantly, a perturbation with wavevector purely in the $xy$-plane still does not elicit an oscillatory response since, due to incompressibility, $\dot{u}_l$ vanishes at this order in wavenumber. In Fig.~\ref{fig:polar_wave_3d}D, we show a polar plot for the eigenfrequency of oscillations for arbitrary $\eta_{o1}$ and display the angular range that elicits an oscillatory response for different values of $\eta_{o1}$ in the inset. This oscillation is left-handed for positive $\eta_{o1}$. 

The handedness of the oscillation for a general wavevector depends on both odd viscosities, $\eta_{o1}$ and $\eta_{o2}$, and on the direction of the mode, through the combination $\nu_{o}$. This changes sign for wavevectors making an angle $\cos^2\theta = (\eta_{o1}-\eta_{o2})/(2\eta_{o1}-\eta_{o2})$, when such a $\theta$ exists. 

Thus, odd viscosity and odd elasticity both lead to oscillatory responses---with distinct characteristics---with wave\emph{number}-independent frequencies thanks to the long-ranged character of Stokesian hydrodynamics. The ratio of viscosity to elasticity is a timescale, which oddness converts from the \emph{relaxation} or \textit{growth} time of a perturbation to the time-period of an oscillation.

\section{Achiral, apolar or both} \label{sec:simpler}

We now examine the response of higher-symmetry columnar material to perturbations, starting with systems which are neither polar nor chiral, ruling out terms that break ${\bf P}\to -{\bf P}$ symmetry or contain an odd number of Levi-Civita tensors. In this limit $\bar{\zeta}_{pc}$ in \eqref{strs1}, $\zeta_{pc}$ in \eqref{elF} and odd viscosities vanish, and \begin{gather}
	\label{achid1}
	\dot{u}_l=-\frac{\cos^2\theta}{\eta}\left[{(2\bar{\mu}+\bar{\lambda})\sin^2\theta}+{\zeta}\right]u_l, \\
	\label{achid2}
	\dot{u}_t=-\frac{1}{\eta}\left[{\bar{\mu}\sin^2\theta}+{\zeta}\right]u_t, 
\end{gather}
are linearly decoupled. Activity, entering solely through $\zeta$, still ensures that the static structure factor of displacement fluctuations, calculated by augmenting \eqref{achid1} and \eqref{achid2} with appropriate white, Gaussian noises \cite{supp}, scales as $\sim 1/q^2$ (for $\zeta>0$) along \emph{all} spatial directions (thereby cutting off the viscosity divergences that arise \cite{ramaswamy1983breakdown} in the equilibrium columnar phase \footnote{Similarly, active lamellar materials \cite{Tap_smec, SJ_chiral_layered, Tap_chol,Ano_sol, Chen_Toner} escape the viscosity divergences of their equilibrium counterparts \cite{Ramaswamy_Toner_Mazenko}}). 

\begin{figure}[t]
	\centering
	\includegraphics[width=0.98\linewidth]{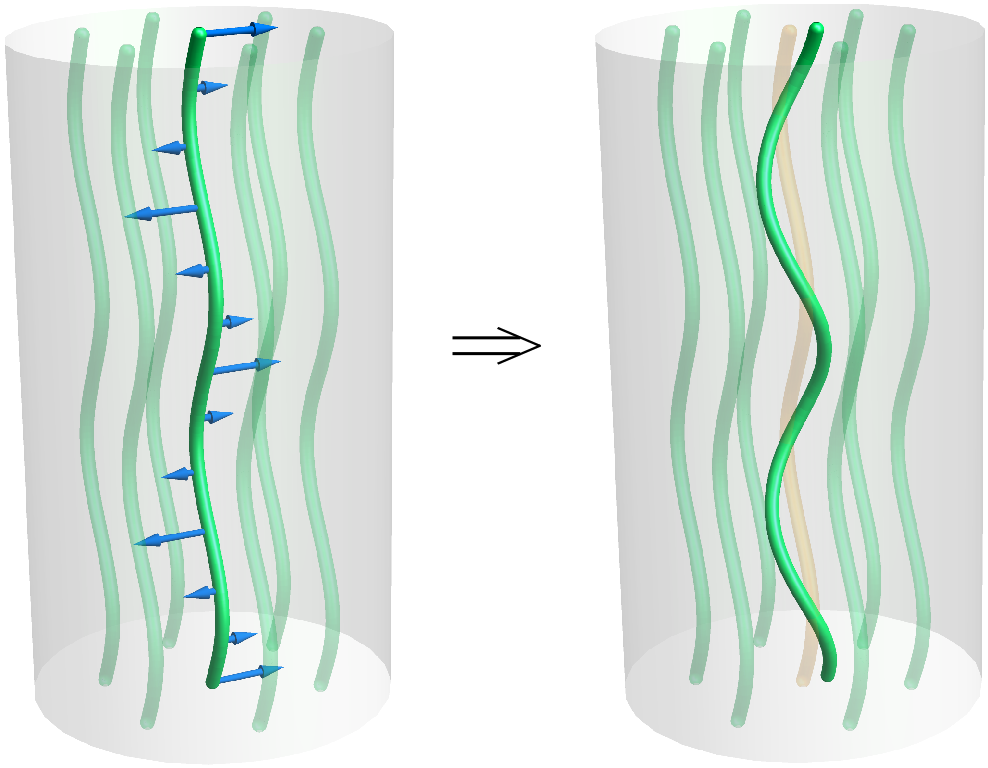}
	\caption{The chiral activity $\zeta_c$ creates transverse flows (blue arrows) to a planar column undulation (green cylinders) and resulting a helical twisting of the columns. For clarity, this is shown only for the highlighted central column, whose initial planar undulation is displayed in orange. The helices are left-handed when $\zeta_c$ is positive and give a macroscopic signature of the microscopic chiral activity.} 
	\label{oddflows}
\end{figure}

We now discuss an apolar but chiral material. All the chiral effects discussed till this point also require breaking inversion symmetry. However, chiral but apolar fluids have extra active stresses which were not included in \eqref{strs1} because they are subdominant to the stress $\propto\bar{\zeta}_{pc}$ in polar materials. In a gradient expansion, the leading-order active chiral stress in apolar materials is 
$\sigma^{ac}_{ij}\propto [\epsilon_{ilk}\partial_l(\partial_k\psi\partial_j\psi)]^S$, introduced in \cite{SJ_chiral_layered}. The resulting force density $\zeta_c\nabla^2\nabla\times{\bf u}_\perp$ \cite{supp} couples $u_l$ and $u_t$ fluctuations, 
\begin{gather}
	\dot{u}_l=-\frac{\cos^2\theta}{\eta}\left[{(2\bar{\mu}+\bar{\lambda})\sin^2\theta}+{\zeta}\right]u_l+\frac{iq\zeta_c\cos\theta}{\eta}u_t, \\
	\dot{u}_t=-\frac{1}{\eta}\left[{\bar{\mu}\sin^2\theta}+{\zeta}\right]u_t-\frac{iq\zeta_c\cos\theta}{\eta}u_l, 
\end{gather}
albeit at higher gradient order than elasticity, yielding, for wavevectors purely along $\hat{{\bf z}}$, mode frequencies $\omega_\pm=-({i}/{\eta})[\zeta\pm\zeta_c q_z]$. That is, chiral activity reduces the relaxation rate of one of the modes while enhancing that of the other. Another example of such a macroscopic manifestation of chirality in an active hydrodynamic instability is the preferential direction of self-shearing in epithelia \cite{Carles}. For $\zeta_c>0$ the favoured mode is a left-handed helical distortion, Fig.~\ref{oddflows}. Interestingly, unlike odd elasticity in two-dimensional chiral solids~\cite{Deboo2} and polar columnar materials discussed above, $\zeta_c$ here resembles a \emph{dissipative} (though of course potentially destabilizing) Onsager coupling between $u_t$ and $u_l$. Note that this coupling originates from the same stress in active model H$^*$ that led to an odd elastic force density in active layered materials directed primarily along contours of constant mean curvature of the layers~\cite{SJ_chiral_layered}. For the columnar phase, in contrast, these transverse chiral flows generate a deformation of the columns into helices even at linear order (see Fig.~\ref{oddflows}). 

Finally, activity fails to create a difference between apolar materials and \textit{achiral} but polar columnar materials in the hydrodynamic limit. The lowest-order additional polar active force density $\zeta_p\partial_z\nabla^2{\bf u}_\perp$ is subleading in gradients, doesn't couple $u_l$ and $u_t$, and simply modifies $\zeta \to \zeta+i\zeta_p q_z$ in \eqref{achid1}, \eqref{achid2}.

\section{Discussion}

We have described the Stokesian hydrodynamics of active columnar phases, including both chiral and polar materials. The polar chiral activity realises an oscillatory response of the columns with a frequency that scales with wavenumber $q$ as $q^0$ and is a ratio of either odd elasticity to even viscosity, or even elasticity to odd viscosity. Estimates for biological tissues using measurements on MDCK epithelial monolayers~\cite{charras2012} suggest a frequency in the $10^{-3}-10^{0}$~Hz range, consistent with the timescale of odd dynamics in Ref.~\cite{Mietke}. Three-dimensionality and the viscous hydrodynamic interaction are essential to the mechanism. Active columnar phases offer an idealised representation of the odd mechanics of many living and synthetic active chiral materials. Axons and epithelia, for example, have a columnar structure and a natural polarity, so chiral activity in these tissues should generate odd dynamical effects. Columnar liquid crystals with macroscopic polarity \cite{polar_columnar_exp}, if suffused with chiral microswimmers, could realise a material in which to test our predictions. The odd dynamics of muscle tissue \cite{shankar2022active} should display distinctive contributions due to chirality.  

The achiral active stress is analogous to an applied mechanical stress and produces a Helfrich-Hurault instability of the columns with degeneracy in the polarisation of the undulations. In chiral active columnar materials, $\zeta_c$ explicitly breaks parity, lifting the degeneracy and favouring one sign of helical column undulation. This highlights the emergence of qualitatively different effects from the same chiral stress in distinct broken-symmetry phases of active model H$^*$. 

An interesting extension of our results would be to determine the effects of confinement and boundary conditions on the column undulations and their associated odd mechanics. Similarly, it will be important to determine the effect of activity on the behaviour of defects \cite{Jacques_lockin} in columnar phases. Finally, our predictions are based largely on linear stability analysis. An understanding of the evolution beyond the linear regime and the final state of the system requires a direct numerical solution of the nonlinear equations of motion. 
\begin{acknowledgments}{We thank Jacques Prost for valuable discussions. SJK acknowledges support through a Raman-Charpak Fellowship of CEFIPRA, hosted by Cesare Nardini, SPEC, CEA-Saclay. AM was supported in part by a TALENT fellowship awarded by CY Cergy Paris Universit\'e and an ANR grant, PSAM. SR was supported in part by a J C Bose Fellowship of the SERB, India and a Simons Visiting Professorship of ICTS. SJK, AM and SR would like to thank the Isaac Newton Institute for Mathematical Sciences, Cambridge, for support and hospitality during the programme New Statistical Physics in Living Matter: non-equilibrium states under Adaptive Control where a part of the work on this paper was undertaken. This work was supported by EPSRC grant no EP/R014604/1. AM and SR thank NORDITA (The Nordic Institute for Theoretical Physics), Sweden for support and hospitality during the programme ``Hydrodynamics at all scales'' where a part of the work on this article was undertaken. AM, GPA and SR thank ICTS-TIFR for support and hospitality during the programme ``Active Matter and Beyond'' during which a part of this work was undertaken.}
\end{acknowledgments}
\bibliography{ref}

\clearpage
\title{Chirality and odd mechanics in active columnar phases\\ Supplementary Material}

%
%
%
%

\maketitle
\onecolumngrid
\appendix
\setcounter{equation}{0}
\section{Recapitulation of chiral and polar active model H}
 A chiral and polar active model H consists of the coupled equations describing a bulk, three-dimensional suspension of polar and chiral elements whose density is denoted by $\psi({\bf r}, t)$, velocity ${\bf v}({\bf r}, t)$ and the degree of polar orientation is accounted for by the polar order parameter ${\bf P}({\bf r},t)$, where ${\bf r}=(x,y,z)$.
The dynamical equation for a conserved scalar field $\psi$ is given by 
\begin{equation}
	\label{psieq3d1_supp}
	\partial_t\psi=-\nabla\cdot({\psi}{\bf v})+M\nabla^2\frac{\delta {F}}{\delta\psi}+\nabla\cdot(\psi{\bf V}_{a})
	+{\xi}_\psi,
\end{equation}
where
\begin{equation}
	\label{PPvel}
	{\bf V}_{a}=\kappa_1(\nabla\psi)\nabla^2\psi+\kappa_2\nabla(\psi\nabla^2\psi)+v_p{\bf P},
\end{equation}
accounts for the active current ignored in the main text (because they turn out to not affect the hydrodynamics of the columnar phase) and $\langle\xi_\psi({\bf r}, t)\xi_\psi({\bf r}', t')\rangle=-2D\nabla^2\delta({\bf r}-{\bf r}')\delta(t-t')$ is a conserving noise also ignored in the main text, which only discussed the deterministic dynamics. The free energy $F$ will be discussed in the next sections.
The velocity field, in the limit of a small Reynolds number relevant for most biological and soft matter systems, has an overdamped, Stokesian dynamics:
\begin{equation}
	\label{vel3dpsi1_supp}
	\eta_{ijkl}\partial_j\partial_kv_l=\psi\partial_i\frac{\delta F[\psi]}{\delta \psi}+\partial_i \Pi-\partial_j\sigma_{ij}^a
	+\xi_{v_i},
\end{equation}
where $\Pi$ is the pressure that enforces the incompressibility constraint $\nabla\cdot{\bf v}=0$, we will consider only one even viscosity in $\eta_{ijkl}$ and $\langle{{\xi}}_{v_i}({\bf r},t){{\xi}}_{v_j}({\bf r}',t')\rangle=-2\delta_{ij}{D}_v{\nabla}^2\delta({\bf r}-{\bf r}')\delta(t-t')$. The active stress $\boldsymbol{\sigma}^a$ and the viscosity tensor $\boldsymbol{\eta}$ will be discussed in the next sections.
In equilibrium, the noise strengths are related via ${D}/{M}={D}_v/{\eta}$, where $\eta$ is the even viscous coefficient. Finally, the equation of motion for the polar order parameter is 
\begin{equation}
	\label{poleq_supp}
	\partial_t{P}_i +(\lambda_PP_j+v_j)\partial_jP_i 
	-({\Omega}_{ij}-\lambda {A}_{ij}) {P}_j=-\Gamma_P\frac{\delta F}{\delta { P}_i}+\xi_{P_i},
\end{equation} 
where advective and self-advective terms which will turn out to be irrelevant for the dynamics of the columnar phase have been explicitly retained. Here, $\Omega_{ij}=(1/2)(\partial_j v_i-\partial_i v_j)$ and $A_{ij}=(1/2)(\partial_j v_i+\partial_i v_j)$. $\xi_{P_i}$ is a Gaussian white noise with a variance $D_P$.

\section{Columnar Density Wave and Active Stresses}
We describe the columnar order by the spatially modulated part $\psi_1$ of a number density field coupled to a polar order parameter ${\bf P}$ that orients the columns with an equilibrium free energy 
\begin{equation}
    F = \int \left\{\frac{\alpha}{2} \psi_1^2 + \frac{\beta}{4} \psi_1^4 + \frac{C_{\parallel}}{2} \bigl( \hat{{\bf P}} \cdot \nabla \psi_1 \bigr)^2 + \frac{C_{\perp}}{2} \bigl[ ({\bf I} - \hat{{\bf P}} \hat{{\bf P}}): \nabla \nabla \psi_1 + q_s^2 \psi_1 \bigr]^2 + \frac{\alpha_P}{2} |{\bf P}|^2 + \frac{\beta_P}{4} |{\bf P}|^4 + \frac{K_{P}}{2} \bigl| \nabla {\bf P} \bigr|^2 \right\}dV .
    \label{eq:SI_free_energy}
\end{equation}
The couplings between ${\bf P}$ and gradients of $\psi$ suppress gradients parallel to ${\bf P}$ and produce Brazovskii-like ordering at wavenumber $q_s$ in directions orthogonal to ${\bf P}$; the operator ${\bf I} - \hat{{\bf P}} \hat{{\bf P}}$ is the projector onto the plane orthogonal to ${\bf P}$. A free energy of the same structure also describes an apolar columnar phase with the liquid crystal director replacing the polar order parameter ${\bf P}$. Polar columnar ground states are obtained when both $\alpha_P<0$ and $\alpha<0$ with a uniform polar order parameter $\hat{{\bf P}} = \hat{{\bf z}}$ and a spatially-modulated density field  
\begin{equation}
    \bar{\psi}_1 = \sum_{{\bf G} \in \Lambda^*} \psi_{1,{\bf G}} \,\mathrm{e}^{i {\bf G} \cdot {\bf x}} ,
\end{equation}
where $\Lambda^*$ is a reciprocal lattice in the $xy$-plane orthogonal to ${\bf P}$. We consider only hexagonal lattices and limit the Fourier series to the fundamental star, explicitly 
\begin{equation}
    {\bf G} \in \biggl\{ \pm q_s \hat{{\bf x}} , \pm q_s \,\frac{- \hat{{\bf x}} + \sqrt{3} \,\hat{{\bf y}}}{2} , \pm q_s \,\frac{\hat{{\bf x}} + \sqrt{3} \,\hat{{\bf y}}}{2} \biggr\} .
\end{equation}
By symmetry, the amplitudes $\psi_{1,{\bf G}}$ all have the same magnitude and we take them to be real, $\psi_{1,{\bf G}} = \psi_1^0$ for all ${\bf G}$. 

Hydrodynamic fluctuations of the columnar state are obtained by introducing an Eulerian displacement field ${\bf u}_{\perp}$, having components only in the $xy$-plane orthogonal to $\hat{{\bf z}}$, into the density modulation 
\begin{equation}
    \psi_1 = \sum_{{\bf G}} \psi_{0} \,\mathrm{e}^{i {\bf G} \cdot ({\bf x} - {\bf u}_{\perp})} ,
\end{equation}
coupled with variations in the direction of polar alignment, $\hat{{\bf P}} = \hat{{\bf z}} + \delta {\bf P}_{\perp}$. The free energy~\eqref{eq:SI_free_energy} then reproduces the usual elasticity of columnar phases \cite{deGen} which reads
\begin{equation}
	F_{{\bf u}_\perp,\delta{\bf P}_\perp}=\int \frac{1}{2}\left[\lambda\text{Tr}[{\bsf E}]^2+2\mu{\bsf E}:{\bsf E}+K\nabla^2u_k\nabla^2u_k +C(\delta{\bf P}_\perp-\partial_z{\bf u}_\perp)^2\right]dV
\end{equation}
where $C=3 C_\parallel{\psi^0_1}^2q_s^2$, $\lambda=3{C}_\perp{\psi^0_1}^2q_s^4$, $\mu=3{C}_\perp{\psi^0_1}^2q_s^4$, $K=3{C}_\perp{\psi^0_1}^2q_s^2$ and we have defined a rotation-invariant strain tensor
\begin{multline}
	E_{ij}=\frac{1}{2}[\partial_iu_j+\partial_ju_i-\partial_iu_k\partial_ju_k-\partial_zu_i\partial_zu_j]\\=\frac{1}{2}\begin{pmatrix}
		2\partial_x u_x-(\partial_x u_x)^2-(\partial_x u_y)^2-(\partial_z u_x)^2 & \partial_x u_y+\partial_y u_x-\partial_x u_x\partial_y u_x-\partial_xu_y\partial_y u_y-\partial_zu_x\partial_zu_y\\\partial_x u_y+\partial_y u_x-\partial_x u_x\partial_y u_x-\partial_xu_y\partial_y u_y-\partial_zu_x\partial_zu_y &2\partial_y u_y-(\partial_y u_x)^2-(\partial_y u_y)^2-(\partial_z u_y)^2
	\end{pmatrix}.
\end{multline}
The suppression of $\psi$ gradients parallel to ${\bf P}$ leads to a coupling $\propto |\delta {\bf P}_{\perp} - \partial_z {\bf u}_{\perp}|^2$, which expresses that pure tilts of the columns do not elicit elastic stresses at linear order. This implies that polarisation fluctuations relax to those determined by the displacement field in a microscopic time (at least for small activities). Since the polarisation fluctuations are not hydrodynamic and relax fast to values determined by displacement fluctuations, the advective and self-advective terms in \eqref{poleq_supp} yield terms at higher order in gradients than we retain and are irrelevant for the hydrodynamics of the polar columnar phase.

The dynamics of the Eulerian displacement field is obtained from \eqref{psieq3d1_supp}
\begin{equation}
	\partial_t{\bf u}_\perp={\bf v}_\perp-{\bf v}\cdot\nabla{\bf u}_\perp-\Gamma\frac{\delta F_{{\bf u}}}{\delta{\bf u}_\perp}+{{\bf V}_{a}}_\perp+\boldsymbol{\xi}_u,
\end{equation}
where $\boldsymbol{\xi}_u$ is a white, Gaussian noise, $\Gamma\propto Mq_s^2$ and ${{\bf V}_{a}}_\perp$ is the active permeation which, for simplicity, we only write to linear order:
\begin{equation}
	{{\bf V}_{a}}_\perp=\mu_1\nabla^2{\bf u}_\perp+\mu_2\nabla_\perp(\nabla_\perp\cdot{\bf u}_\perp),
\end{equation}
where $\mu_1$ and $\mu_2$ depend on $\kappa_{1,2}$ and $v_p$. However, these  terms at $\mathcal{O}(\nabla^2)$ are subdominant to the terms that appear through ${\bf v}_\perp$ due to Stokesian dynamics, and therefore, we do not consider them further.

We focus now on the hydrodynamic form of the active contributions to the stress. We consider four distinct active stresses: an apolar achiral stress proportional to $\nabla\psi \nabla \psi$; a similar term proportional to ${\bf PP}$; a polar chiral stress proportional to the symmetric part of $({\bf P} \times \nabla \psi) \,\nabla \psi$; and an apolar chiral stress proportional to the symmetric part of $\nabla \times (\nabla \psi \nabla \psi)$ (which we discuss later). 
We treat the term proportional to ${\bf PP}$ first: its linearisation is 
\begin{equation}
    {\bf PP} = P_0^2 \bigl( \hat{{\bf z}} \hat{{\bf z}} + \hat{{\bf z}} \,\delta {\bf P}_{\perp} + \delta {\bf P}_{\perp} \,\hat{{\bf z}} \bigr) = P_0^2 \bigl( \hat{{\bf z}} \hat{{\bf z}} + \hat{{\bf z}} \,\partial_z {\bf u}_{\perp} + \partial_z {\bf u}_{\perp} \,\hat{{\bf z}} \bigr) ,
    \label{eq:SI_PPstress}
\end{equation}
where $P_0 = \sqrt{-\alpha_p/\beta_P}$ is the preferred magnitude of ${\bf P}$. 
The analysis of the stresses involving $\psi$ all follow from the hydrodynamic part of $\nabla\psi \nabla\psi$, which is 
\begin{align}
    \nabla \psi \nabla \psi & = \sum_{{\bf G}_1 , {\bf G}_2} \psi_0^2 \bigl[ ({\bf I} - \nabla {\bf u}_{\perp}) \cdot i {\bf G}_1 \bigr] \bigl[ ({\bf I} - \nabla {\bf u}_{\perp}) \cdot i {\bf G}_2 \bigr] \,\mathrm{e}^{i ({\bf G}_1 + {\bf G}_2) \cdot ({\bf x} - {\bf u}_{\perp})} , \\
    & = \sum_{{\bf G}} \psi_0^2 \bigl( {\bf I} - \nabla {\bf u}_{\perp} \bigr) \cdot {\bf GG} \cdot \bigl( {\bf I} - (\nabla {\bf u}_{\perp})^T \bigr) + \cdots ,
\end{align}
where we retain explicitly only the terms with ${\bf G}_1 + {\bf G}_2 = 0$. For the fundamental star of the hexagonal lattice 
\begin{equation}
    \sum_{{\bf G}} {\bf GG} = 3q_s^2 \bigl( \hat{{\bf x}} \hat{{\bf x}} + \hat{{\bf y}} \hat{{\bf y}} \bigr) ,
\end{equation}
so that to linear order (and retaining only the hydrodynamic part)
\begin{equation}
    \nabla \psi \nabla \psi = 3 q_s^2 \psi_0^2 \bigl( \hat{{\bf x}} \hat{{\bf x}} + \hat{{\bf y}} \hat{{\bf y}} - \nabla {\bf u}_{\perp} - (\nabla {\bf u}_{\perp})^T \bigr) .
    \label{eq:SI_gradpsi_gradpsi_stress}
\end{equation}
The full, nonlinear expression for $\nabla \psi \nabla \psi$ is 
\begin{equation}
	\label{nlinstr}
	\nabla \psi \nabla \psi =3{\psi_1^0}^2q_s^2\begin{pmatrix}(\partial_x u_x-1)^2+(\partial_x u_y)^2 & \partial_yu_x(\partial_ xu_x-1)+(\partial_yu_y-1)\partial_xu_y & \partial_zu_x(\partial_x u_x-1)+\partial_zu_y\partial_xu_y\\\partial_yu_x(\partial_xu_x-1)+\partial_xu_y(\partial_yu_y-1) &(\partial_y u_x)^2+(\partial_yu_y-1)^2 &\partial_zu_x\partial_yu_x+\partial_zu_y(\partial_yu_y-1)\\\partial_zu_x(\partial_xu_x-1)+\partial_zu_y\partial_xu_y&\partial_zu_x\partial_yu_x+\partial_zu_y(\partial_yu_y-1) &(\partial_z u_x)^2+(\partial_zu_y)^2\end{pmatrix}.
\end{equation}
For the polar chiral contribution a direct calculation gives 
\begin{equation}
    \begin{split}
        \bigl( {\bf P} \times \nabla \psi \bigr) \nabla \psi = 3q_s^2 P_0 \psi_0^2 & \Bigl[ \hat{{\bf y}} \hat{{\bf x}} - \hat{{\bf x}} \hat{{\bf y}} + \bigl( \partial_x u_y + \partial_y u_x \bigr) \bigl( \hat{{\bf x}} \hat{{\bf x}} - \hat{{\bf y}} \hat{{\bf y}} \bigr) + \bigl( \partial_y u_y - \partial_x u_x \bigr) \bigl( \hat{{\bf x}} \hat{{\bf y}} + \hat{{\bf y}} \hat{{\bf x}} \bigr) \\
        & \quad + \bigl( \partial_x u_x + \partial_y u_y \bigr) \bigl( \hat{{\bf x}} \hat{{\bf y}} - \hat{{\bf y}} \hat{{\bf x}} \bigr) + \partial_z u_x \bigl( \hat{{\bf z}} \hat{{\bf y}} - \hat{{\bf y}} \hat{{\bf z}} \bigr) + \partial_z u_y \bigl( \hat{{\bf x}} \hat{{\bf z}} - \hat{{\bf z}} \hat{{\bf x}} \bigr) \Bigr] ,
    \end{split}
\end{equation}
retaining only the linear hydrodynamic terms. 

Combining these results we see that the active stress 
\begin{equation}
    \boldsymbol{\sigma}^a = - \zeta_{H} \nabla\psi \nabla\psi + \zeta_{pa} \,{\bf PP} + \bar{\zeta}_{pc} \bigl[ ({\bf P}\times\nabla\psi) \nabla\psi \bigr]^S ,
\end{equation}
where ${\bf M}^S$ denotes the symmetrisation of the tensor ${\bf M}$, implies a hydrodynamic active force density 
\begin{equation}
    \begin{split}
        \nabla \cdot \boldsymbol{\sigma}^a & = 3q_s^2 \psi_0^2 \zeta_{H} \bigl( \nabla^2 {\bf u} + \nabla (\nabla_{\perp} \cdot {\bf u}_{\perp}) \bigr) + \zeta_{pa} P_0^2 \bigl( \partial^2_{z} {\bf u}_{\perp} + \hat{{\bf z}} \,\partial_z (\nabla_{\perp} \cdot {\bf u}_{\perp}) \bigr) + 3 q_s^2 P_0 \psi_0^2 \bar{\zeta}_{pc} \nabla_{\perp}^2 \bigl( u_y \,\hat{{\bf x}} - u_x \,\hat{{\bf y}} \bigr) .
    \end{split}
\end{equation}
This can be rewritten as
\begin{equation}
	\begin{split}
		\nabla \cdot \boldsymbol{\sigma}^a & = (3q_s^2 \psi_0^2 \zeta_{H} +\zeta_{pa} P_0^2)\nabla^2 {\bf u} - \zeta_{pa} P_0^2 \bigl[ \nabla_\perp^2 {\bf u}_{\perp} +\nabla_\perp (\nabla_{\perp} \cdot {\bf u}_{\perp}) \bigr] + 3 q_s^2 P_0 \psi_0^2 \bar{\zeta}_{pc} \nabla_{\perp}^2 \bigl( u_y \,\hat{{\bf x}} - u_x \,\hat{{\bf y}} \bigr) .
	\end{split}
\end{equation}
upon absorbing total gradients arising from $\zeta_H$ and $\zeta_{pa}$ into a redefinition of the pressure.
After combining with the passive elasticity, this gives the elastic force density $\boldsymbol{{\cal F}}^e$ of Eq.~$(8)$ in the main text with the definitions $\bar{\mu}=\mu-\zeta_1$ and $\bar{\lambda}=\lambda-\zeta_2$ and $\zeta_1=\zeta_{pa} P_0^2 $, $\zeta_2=0$, $\zeta=(3q_s^2 \psi_0^2 \zeta_{H} +\zeta_{pa} P_0^2)$ and $\zeta_{pc}=3 q_s^2 P_0 \psi_0^2 \bar{\zeta}_{pc}$. We retain $\zeta_2$ (which is $0$ in our description) to remind the reader that, in general, the bulk modulus is renormalised by activity (for instance, if one retains an anisotropic version of the $\nabla\psi\nabla\psi$ stress).

While in this article, we confine ourselves to a discussion of the linear physics of active columnar materials, note that our description that starts from an active model H automatically yields the fully covariant nonlinear equations of motion for the active columnar phase (as we show here) which should serve as the starting point for a numerical simulation of such materials.

\section{Odd viscosity in polar and chiral fluids}
In equilibrium polar fluids, with at least six-fold symmetry transverse to the polar direction, the viscosity tensor has the following form:
\begin{equation}
	\eta_{ijkl}=\eta_{1}P_iP_jP_kP_l+{\eta}(\delta_{ik}\delta_{jl}+\delta_{il}\delta_{jk})+\frac{\eta_3}{4}(P_iP_k\delta_{jl}+P_jP_k\delta_{il}+P_iP_l\delta_{jk}+P_jP_l\delta_{ik})+\eta_4\delta_{ij}\delta_{kl}+\eta_5(\delta_{ij}P_kP_l+P_iP_j\delta_{kl}).
\end{equation}
The last two vanish in incompressible systems leaving a viscous stress tensor
\begin{equation}
	{\sigma_{ij}}^v_{u}=\eta_{1}P_iP_jP_kP_lA_{kl}+2\eta A_{ij}+\frac{\eta_3}{2}(P_iP_kA_{jk}+P_jP_kA_{ik}).
\end{equation}
Of these, we only retain $\eta$ in our discussion in the main text and will do so here as well. 
In chiral and polar materials, one has additional velocity-dependent active stresses at the same order in gradients
\begin{equation}
	\label{chivis_supp}
	{\sigma_{ij}}^v_o=2\left[\epsilon_{ilk}P_k\left\{2{{\eta}_{o1}}P_jP_mA_{lm}+{\eta}_{o2}\left(A_{lj}-P_jP_mA_{lm}\right)\right\}\right]^S.
\end{equation}
The divergence of the viscous stresses $\partial_j({\sigma_{ij}}^v_o+{\sigma_{ij}}^v_{u})=\eta_{ijkl}\partial_j\partial_kv_l$ gives the L.H.S. of \eqref{vel3dpsi1} (with $\eta_1=\eta_3=0$). In particular, the linearised version of \eqref{chivis_supp}
\begin{equation}
	\boldsymbol{{\cal F}}_{o}^{v} = \eta_{o1} \bigl( \partial_{zz} \boldsymbol{\epsilon}\cdot{\bf v}_{\perp} + \boldsymbol{\epsilon} \cdot \nabla_{\perp} \partial_z v_z + \partial_z (\partial_x v_y - \partial_y v_x) \,\hat{{\bf z}} \bigr) + \eta_{o2} \nabla_{\perp}^2 \boldsymbol{\epsilon}\cdot{\bf v}_{\perp} ,  
\end{equation}
is the odd viscous force density (Eq.~$(10)$ of the main text).

\section{Linear Hydrodynamics of Active Columnar Phases}
The hydrodynamic variables in the columnar phase are the Eulerian displacement field ${\bf u}_{\perp}$ of the columns and the fluid velocity ${\bf v}$. Neglecting permeation, the linear hydrodynamics of the displacement field is 
\begin{equation}
    \partial_t {\bf u}_{\perp} = {\bf v}_{\perp} ,
\end{equation}
while that for the fluid velocity is the Stokes equation (Eq.~$(11)$ of the main text)
\begin{equation}
    0 = - \nabla \Pi + \eta \nabla^2 {\bf v} + \boldsymbol{{\cal F}}_{o}^{v} + \boldsymbol{{\cal F}}^{e} ,
\end{equation}
together with incompressibility $\nabla \cdot {\bf v} = 0$. Here, $\Pi$ is the pressure
and $\boldsymbol{{\cal F}}^e$ is the elastic force density (Eq.~$(8)$ of the main text)
\begin{equation}
    \boldsymbol{{\cal F}}^{e} = \bar{\mu} \nabla_{\perp}^2 {\bf u}_{\perp} + (\bar{\mu}+\bar{\lambda}) \nabla_{\perp} \nabla_{\perp} \cdot {\bf u}_{\perp} + \zeta \nabla^2 {\bf u}_{\perp} + \zeta_{pc} \nabla_{\perp}^2 \boldsymbol{\epsilon}\cdot{\bf u}_{\perp} . 
\end{equation}

We solve the Stokes equation for the velocity in terms of the displacement field by Fourier transform 
\begin{equation}
    0 = - i \,{\bf q} \,\tilde{\Pi} - \eta q^2 \,\tilde{{\bf v}} - \eta_{o1} \Bigl( q_z^2 \,\boldsymbol{\epsilon} + q_z \bigl[ (\boldsymbol{\epsilon}\cdot{\bf q}_{\perp}) \hat{{\bf z}} - \hat{{\bf z}} (\boldsymbol{\epsilon}\cdot{\bf q}_{\perp}) \bigr] \Bigr) \cdot \tilde{{\bf v}} - \eta_{o2} q_{\perp}^2 \,\boldsymbol{\epsilon}\cdot\tilde{{\bf v}}_{\perp} + {\bf F}_{\bf q}
\end{equation}
In developing the solution we make use of the natural decomposition of the vector space $\mathbb{R}^3$ as $\mathbb{R}^2\oplus\mathbb{R}$. There are two such splittings: One comes from the columnar structure, where the (polar) orientation of the columns defines the $\mathbb{R}$ factor, which we take to be $\hat{{\bf z}}$, and the vector $\tilde{{\bf u}}_{\perp}$ lies entirely in the two-dimensional subspace orthogonal to this ($xy$-plane). The other comes from the wavevector ${\bf q}$, whose direction also provides a splitting $\mathbb{R}^3 \cong \mathbb{R}^2\oplus\mathbb{R}$, and as the flow is incompressible, ${\bf q} \cdot \tilde{{\bf v}} = 0$, the velocity $\tilde{{\bf v}}$ lies entirely in the orthogonal two-dimensional subspace. Since our objective here is to solve for $\tilde{{\bf v}}$ as a function of $\tilde{{\bf u}}$, we make use of the latter splitting. 

Let $\{ {\bf e}_1, {\bf e}_2, {\bf e}_3\}$ be an orthonormal frame with ${\bf q} = q \,{\bf e}_3$ and $\hat{{\bf z}}$ lying in the ${\bf e}_1,{\bf e}_3$-plane ({\sl i.e.} ${\bf e}_2$ directs the intersection of the plane orthogonal to ${\bf q}$ with the $xy$-plane). This amounts to 
\begin{align}
    & \hat{{\bf z}} = \frac{q_z}{q} \,{\bf e}_3 - \frac{q_{\perp}}{q} \,{\bf e}_1 , && {\bf q}_{\perp} = \frac{q_{\perp}^2}{q} \,{\bf e}_3 + \frac{q_z q_{\perp}}{q} \,{\bf e}_1 , && \boldsymbol{\epsilon}\cdot{\bf q}_{\perp} = -q_{\perp} \,{\bf e}_2 .
\end{align}
In this basis the Stokes equation reads 
\begin{equation}
\begin{split}
    0 & = - \biggl[ \eta q^2 \bigl( {\bf e}_1 {\bf e}_1 + {\bf e}_2 {\bf e}_2 \bigr) - \bigl[ \eta_{o1} (q_z^2-q_{\perp}^2) + \eta_{o2} q_{\perp}^2 \bigr] \frac{q_z}{q} \bigl( {\bf e}_2 {\bf e}_1 - {\bf e}_1 {\bf e}_2 \bigr) \biggr] \cdot \tilde{{\bf v}} + \bigl( {\bf e}_1 {\bf e}_1 + {\bf e}_2 {\bf e}_2 \bigr) \cdot {\bf F}_{\bf q} \\
    & \quad + {\bf e}_3 \biggl[ - i q \,\tilde{\Pi} - \frac{(2 \eta_{o1} q_z^2 + \eta_{o2} q_{\perp}^2) q_{\perp}}{q} \,{\bf e}_2 \cdot \tilde{{\bf v}} + {\bf e}_3 \cdot {\bf F}_{\bf q} \biggr] ,
\end{split}
\end{equation}
The component parallel to ${\bf q}$ gives an expression for the pressure 
\begin{equation}
i \tilde{\Pi} = - \frac{(2 \eta_{o1} q_z^2 + \eta_{o2} q_{\perp}^2) q_{\perp}}{q^2} \,{\bf e}_{2} \cdot \tilde{{\bf v}} + \frac{1}{q} \,{\bf e}_3 \cdot {\bf F}_{\bf q} ,
\end{equation}
while the orthogonal components give the fluid velocity as 
\begin{equation}
\tilde{{\bf v}} = \biggl[ \frac{\eta}{\Delta q^2} \bigl( {\bf e}_1 {\bf e}_1 + {\bf e}_2 {\bf e}_2 \bigr) + \frac{\nu_{o} q_z}{\Delta q^3} \bigl( {\bf e}_2 {\bf e}_1 - {\bf e}_1 {\bf e}_2 \bigr) \biggr] \cdot {\bf F}_{\bf q} ,
\end{equation}
where 
\begin{gather}
    \nu_{o} = \frac{\eta_{o1} (q_z^2-q_{\perp}^2) + \eta_{o2} q_{\perp}^2}{q^2} , \\
    \Delta = \eta^2 + \frac{\nu_{o}^2 q_z^2}{q^2} .
\end{gather}
The two terms in the inverse viscosity operator are the even and odd parts of the mobility ${\bf M}_{{\bf q}}$. 
It can be seen directly that the odd mobility vanishes if $q_z = 0$. 
It also vanishes along the cone $\nu_{o} = 0$, or 
\begin{equation}
\frac{q_z^2}{q^2} = \frac{\eta_{o1} - \eta_{o2}}{2 \eta_{o1} - \eta_{o2}} , 
\end{equation}
and changes sign between the inside and outside of this cone. 
The consequence of this is that the sense of rotation in the column oscillations can switch as a function of the direction of the wave relative to the column axis.

\section{Displacement Dynamics}

The dynamics of the displacement field, $\partial_t \tilde{{\bf u}}_{\perp} = \tilde{{\bf v}}_{\perp}$, takes place in the $xy$-plane orthogonal to the (polar) column axis and it is convenient to express it using a basis adapted to this splitting. Let $\{ {\bf e}_l , {\bf e}_t , \hat{{\bf z}} \}$ be an orthonormal frame with ${\bf q} = q_z \,\hat{{\bf z}} + q_{\perp} \,{\bf e}_l$. One finds 
\begin{align}
    & {\bf e}_1 = \frac{q_z}{q} \,{\bf e}_{l} - \frac{q_{\perp}}{q} \,\hat{{\bf z}} , 
    && {\bf e}_2 = {\bf e}_t ,
\end{align}
and then the displacement dynamics reads 
\begin{equation}
	\partial_t \tilde{{\bf u}}_{\perp} = \biggl[ \frac{\eta}{\Delta q^2} \biggl( \frac{q_z^2}{q^2} {\bf e}_{l} {\bf e}_{l} + {\bf e}_t {\bf e}_t \biggr) - \frac{\nu_{o} q_z^2}{\Delta q^4} \,\boldsymbol{{\epsilon}} \biggr] \cdot{\bf F}_{\bf q}
\end{equation}
In the same basis the elastic force density is 
\begin{equation}
{\bf F}_{\bf q} = - \biggl[ \bigl( \bar{\mu} q_{\perp}^2 + \zeta q^2 \bigr) \bigl( {\bf e}_{l} {\bf e}_{l} + {\bf e}_t {\bf e}_t \bigr) + (\bar{\mu}+\bar{\lambda}) q_{\perp}^2 \,{\bf e}_l {\bf e}_l + \zeta_{pc} q_{\perp}^2 \boldsymbol{{\epsilon}} \biggr] \cdot \tilde{{\bf u}}_{\perp} .
\end{equation}

The structure of the linear dynamics is $\partial_t \tilde{{\bf u}}_{\perp} = {\bf D}\cdot\tilde{{\bf u}}_{\perp}$, where the `dynamical matrix' ${\bf D}$ can be written (in the ${\bf e}_l , {\bf e}_t$ basis) in the form 
\begin{equation}
	\label{dynmax}
{\bf D} = D_0 \begin{bmatrix} 1 & 0 \\ 0 & 1 \end{bmatrix} + D_{\textrm{r}} \begin{bmatrix} 0 & -1 \\ 1 & 0 \end{bmatrix} + D_{+} \begin{bmatrix} 1 & 0 \\ 0 & -1 \end{bmatrix} + D_{\times} \begin{bmatrix} 0 & 1 \\ 1 & 0 \end{bmatrix} ,
\end{equation}
and the coefficients are
\begin{align}
D_0 & = - \frac{\eta}{2\Delta q^4} \Bigl[ \bigl( \bar{\mu} q_{\perp}^2 + \zeta q^2 \bigr) \bigl( 2q_z^2 + q_{\perp}^2 \bigr) + (\bar{\mu} + \bar{\lambda}) q_z^2 q_{\perp}^2 \Bigr] - \frac{\nu_{o} \zeta_{pc} q_z^2 q_{\perp}^2}{\Delta q^4} , \\
D_{\textrm{r}} & = \frac{\eta \zeta_{pc} (2q_z^2 + q_{\perp}^2) q_{\perp}^2}{2\Delta q^4} - \frac{\nu_{o} q_z^2}{2\Delta q^4} \bigl[ 2 \bigl( \bar{\mu} q_{\perp}^2 + \zeta q^2 \bigr) + (\bar{\mu}+\bar{\lambda}) q_{\perp}^2 \bigr] , \\
D_{+} & = \frac{\eta q_{\perp}^2}{2\Delta q^4} \bigl[ \bar{\mu} q_{\perp}^2 + \zeta q^2 - (\bar{\mu} + \bar{\lambda}) q_z^2 \bigr] , \\
D_{\times} & = \frac{\eta \zeta_{pc} q_{\perp}^4}{2\Delta q^4} - \frac{\nu_{o} (\bar{\mu}+\bar{\lambda}) q_z^2 q_{\perp}^2}{2 \Delta q^4} .
\end{align}
The character of the dynamics depends on the sign of $D_{\textrm{r}}^2 - D_{+}^2 - D_{\times}^2$; when it is positive the dynamics is oscillatory, 
which is the regime we focus on. The eigenmodes in the oscillatory regime are given by 
\begin{equation}
\tilde{{\bf u}} = u_0 \bigl[ (D_{\textrm{r}} - D_{\times}) \,{\bf e}_l + (D_{+} \pm i \omega) \,{\bf e}_t \bigr] \,\mathrm{e}^{D_0 t \mp i \omega t} ,
\end{equation}
where $\omega = \sqrt{D_{\textrm{r}}^2 - D_{+}^2 - D_{\times}^2}$ is the oscillation frequency. Growth or decay of these modes is determined by the sign of $D_0$ and the transition between them marks a Hopf bifurcation. 

\section{Apolar and chiral force density}
 We now discuss the effect of a chiral (but apolar) stress $\bar{z}_c[\nabla \times (\nabla \psi \nabla \psi)]^S$. In terms of the displacement field, the force density associated with this stress is $\zeta_c\nabla^2\nabla\times{\bf u}_\perp$, with $\zeta_c=3\bar{z}_c{\psi^0_1}^2q_s^2$. 
Upon including the effect of this stress, the displacement dynamics in Fourier space becomes 
\begin{equation}
    \partial_t \tilde{{\bf u}}_{\perp} = \biggl[ \frac{\eta}{\Delta q^2} \biggl( \frac{q_z^2}{q^2} {\bf e}_{l} {\bf e}_{l} + {\bf e}_t {\bf e}_t \biggr) - \frac{\nu_{o} q_z^2}{\Delta q^4} \,\boldsymbol{{\epsilon}} \biggr] \cdot \tilde{\boldsymbol{{\cal F}}}^e - \biggl[ \frac{\eta q_z q_{\perp}}{\Delta q^4} \,{\bf e}_l + \frac{\nu_{o} q_z q_{\perp}}{\Delta q^4} \,{\bf e}_t \biggr] \bigl( \hat{{\bf z}} \cdot \tilde{\boldsymbol{{\cal F}}}^e \bigr) .
\end{equation}
where 
\begin{equation}
	 \tilde{\boldsymbol{{\cal F}}}^e=- \biggl[ \bigl( \bar{\mu} q_{\perp}^2 + \zeta q^2 \bigr) \bigl( {\bf e}_{l} {\bf e}_{l} + {\bf e}_t {\bf e}_t \bigr) + (\bar{\mu}+\bar{\lambda}) q_{\perp}^2 \,{\bf e}_l {\bf e}_l + \bigl(\zeta_{pc} q_{\perp}^2 - i \zeta_c q^2 q_z \bigr) \boldsymbol{\epsilon}+ i \zeta_c q^2 q_{\perp} \,\hat{\bf z} {\bf e}_t \biggr] \cdot \tilde{{\bf u}}_{\perp}.
\end{equation}
The dynamical matrix can still be written as in \eqref{dynmax}, but now with the following definitions:
\begin{align}
	D_0 & = - \frac{\eta}{2\Delta q^4} \Bigl[ \bigl( \bar{\mu} q_{\perp}^2 + \zeta q^2 \bigr) \bigl( 2q_z^2 + q_{\perp}^2 \bigr) + (\bar{\mu} + \bar{\lambda}) q_z^2 q_{\perp}^2 \Bigr] - \frac{\nu_{o} [\zeta_{pc} q_z^2 q_{\perp}^2-i\zeta_cq_zq^2(q_z^2+q^2)]}{\Delta q^4} , \\
	D_{\textrm{r}} & = \frac{\eta [\zeta_{pc} (2q_z^2 + q_{\perp}^2) q_{\perp}^2-2i\zeta_cq^4q_z]}{2\Delta q^4} - \frac{\nu_{o} q_z^2}{2\Delta q^4} \bigl[ 2 \bigl( \bar{\mu} q_{\perp}^2 + \zeta q^2 \bigr) + (\bar{\mu}+\bar{\lambda}) q_{\perp}^2 \bigr] , \\
	D_{+} & = \frac{\eta q_{\perp}^2}{2\Delta q^4} \bigl[ \bar{\mu} q_{\perp}^2 + \zeta q^2 - (\bar{\mu} + \bar{\lambda}) q_z^2 \bigr] - \frac{i \zeta_c \nu_o q_z q_{\perp}^2}{2\Delta q^2}, \\
	D_{\times} & = \frac{\eta \zeta_{pc} q_{\perp}^4}{2\Delta q^4} - \frac{\nu_{o} (\bar{\mu}+\bar{\lambda}) q_z^2 q_{\perp}^2}{2 \Delta q^4} .
\end{align}



\end{document}